\documentclass[11pt]{article}
\usepackage{amssymb}
\usepackage{amsmath}
\usepackage{amsthm}

\usepackage{graphicx,color}

\def\openone{\leavevmode\hbox{\small1\kern-3.8pt\normalsize1}}
\def\tr{{\rm Tr}}

\def\cM{{\Lambda}}

\def\tq{{\tilde{q}}}
\def\te{{\tilde{e}}}
\newcommand{\sk}{\mathfrak{s}}

\newtheorem{theorem}{Theorem}

\theoremstyle{definition}
\newtheorem{definition}{Definition}

\def\reff#1{(\ref{#1})}
\def\eps{\varepsilon}

\newcommand{\cB}{{\cal B}}
\newcommand{\cD}{{\cal D}}
\newcommand{\cO}{\mathcal{O}}

\newcommand{\cH}{{\cal H}}

\newcommand{\cP}{{\cal P}}

\DeclareRobustCommand\openone{\leavevmode\hbox{\small1\normalsize\kern-.33em1}}

\newcommand{\be}{\begin{equation}}
\newcommand{\ee}{\end{equation}}
\newcommand{\bea}{\begin{eqnarray}}
\newcommand{\eea}{\end{eqnarray}}
\newcommand{\beas}{\begin{eqnarray*}}
\newcommand{\eeas}{\end{eqnarray*}}

\begin{document}

\title{\LARGE\bf
  An upper bound on the second order asymptotic expansion for the quantum communication cost of state redistribution}
\author{Nilanjana Datta$^1$, Min-Hsiu Hsieh$^2$ and Jonathan Oppenheim$^{3,4}$\\[3mm]
 \small\it $^1$Statistical Laboratory, Centre for Mathematical Sciences, University of Cambridge,\\ \small\it Wilberforce Road, Cambridge CB3 0WA, U.K.\\[1mm]
 \small\it $^2$Centre for Quantum Computation and Intelligent Systems (QCIS),\\ \small\it Faculty of Engineering and Information Technology (FEIT),\\
\small\it University of Technology Sydney (UTS), NSW 2007, Australia\\[1mm]
 \small\it $^3$Department of Physics and Astronomy, University College London,\\ \small\it
Gower Street, London WC1E 6BT, United Kingdom\\[1mm]
 \small\it $^4$ Department of Computer Science and Centre for Quantum Technologies,\\ \small\it
National University of Singapore, Singapore 119615
}

\date{}

\maketitle

\begin{abstract}
State redistribution is the protocol in which, given an arbitrary tripartite quantum state, with two of the subsystems initially being with Alice and one being with Bob,
the goal is for Alice to send one of her subsystems to Bob, possibly with the help of prior shared entanglement.
We derive an upper bound on the second order asymptotic expansion for the quantum communication cost of achieving state redistribution with a given finite accuracy. 
In proving our result, we also obtain an upper bound on the quantum communication cost of this protocol in the one-shot setting, by using the protocol of coherent state merging as a primitive.
\end{abstract}

\section{Introduction}\label{intro}

State redistribution is a fundamental protocol in quantum information theory and
serves as a primitive for various other information theoretic protocols, such as state merging, coherent state merging,
and quantum channel simulation and  rate distortion in the presence of quantum side information~(see for e.g.\cite{DY06, WDHW13} and references therein). It can be described as follows. Suppose Alice and Bob share a tripartite state $\rho_{ABC}$ with the systems $A$ and $C$ being with Alice and the system $B$ being with Bob. Let $\psi_{ABCR}$ denote a purification of $\rho_{ABC}$, with $R$ being the inaccessible, purifying reference system. In addition, Alice and Bob are allowed to share entangled states. The task is for Alice to transfer the state of her system $A$ to Bob, possibly with the help of the prior shared entanglement, such that the purity of the global state
is preserved. Alice and Bob
can both do local operations (LO) on systems in their possession and Alice can send qubits to
Bob, i.e., she is allowed one-way quantum communication (QC) with Bob. The minimum number of qubits needed for this task is referred to as the {\em{quantum communication cost}} of the protocol. 

This protocol was first introduced by Devetak and Luo in~\cite{DL09}. It was studied by Devetak and Yard~\cite{YD09, DY08} in the so-called `asymptotic i.i.d.~setting', in which Alice and Bob share multiple (say $n$) identical copies of the state $\rho_{ABC}$, instead of just one. The quantum communication cost, $Q$, in this setting is defined as the minimum rate 
of quantum communication from Alice to Bob needed so that the error incurred in achieving the goal (of 
transferring the states of the systems labelled by $A$ from Alice to Bob) vanishes in the asymptotic limit ($ n \to \infty$). Let the corresponding rate of entanglement consumption\footnote{A negative entanglement consumption rate implies that entanglement is
instead generated.} be denoted as $E$. Devetak and Yard~\cite{YD09} proved that state redistrubution is possible in this setting if and only if $Q$ and $E$ satisfy the following bounds:
\be\label{asymp} Q \geq \frac{1}{2}  I(A;R|B); \quad Q+E \geq H(A|B).\ee
Here $I(A;R|B)$ denotes the conditional mutual information of the state $\rho_{ABR} := \tr_C\psi_{ABCR}$, and $H(A|B)$ is the conditional entropy of $\rho_{AB}$. In fact, this provided the first
 operational interpretation of the quantum conditional mutual information~\cite{DY06, YD09}.

In this paper, we first consider state redistribution in the `one-shot setting' in which Alice and Bob share a {\em{single copy}} of the state $\rho_{ABC}$. Instead of requiring that the error incurred in the protocol vanishes asymptotically, it is
natural in this case to allow for a small but non-zero error $\eps >0$. 
We refer to the minimum number of qubits which are needed to be transferred 
in this case as the {\em{ $\eps$-error one-shot quantum communication cost}}. We derive an upper bound on this quantity in terms of the smooth min- and max-entropies of One-shot Information Theory~(see e.g.~\cite{renner, marco} and references therein).
%

Our ultimate goal is to derive an upper bound on the {\em{second order asymptotic expansion}} for the quantum communication cost of state redistribution, for $n$ identical copies of the state $\rho_{ABC}$, with an error of at most $\eps$. 
We establish that, for any tripartite state $\rho_{ABC}$, and any given $\eps \in (0,1)$, an upper bound on the quantum communication cost of achieving quantum state
redistribution of $\rho_{ABC}^{\otimes n}$, with an error of at most $\eps$, can be expressed in the form
$$an + b\sqrt{n} + \cO(\log n);$$
here the first order coefficient $a$ is equal to $\frac{1}{2}  I(A;R|B)$ (as expected from the result of Devetak and Yard~\cite{YD09}). We obtain an explicit expression for the second order coefficient $b$, which depends on both the state $\rho_{ABC}$ and the allowed error threshold $\eps$.

A simple corollary of the above expansion is the following result of~\cite{YD09}: in the asymptotic i.i.d. setting, state redistribution can be achieved if Alice sends qubits at a rate $(1/2) I(A;R|B)$ to Bob (as is implied by \reff{asymp}).

Our result employs the protocol of {\em{coherent state merging}}\footnote{This protocol is also known as FQSW, which is an acronym for Fully Quantum Slepian Wolf.}~\cite{FQSW} which is described 
in the one-shot setting as follows. One starts with a tripartite pure state ${\psi_{ABR}}$, where the system $A$ is with Alice, $B$ is with Bob and $R$ denotes the purifying reference system. Alice and Bob do not share any entanglement at  the start of the protocol. The aim is for Alice to transfer the state of the system $A$ to Bob and at the same time generate entanglement with him. Alice and Bob can both do local operations on systems in their possession and Alice can send qubits to Bob. The quantities of interest are the quantum communication cost and the entanglement gain: the former is the minimum number of qubits that Alice needs to send to Bob in order to achieve the state transfer (up to a given finite accuracy) and the latter is the maximum entanglement created in this process. In~\cite{DH11} we obtained bounds on these quantities under the constraint that the error incurred in the protocol was at most $\eps$ (for an arbitrary but fixed $\eps \in (0,1)$).  

It is easy to see that this protocol can be considered as a special case of state redistribution: the system $C$ which Alice has at the start of state redistribution can be viewed as quantum side information; then coherent state merging corresponds to the case in which no such side information is available to Alice. In this sense, state redistribution can be used as a primitive for coherent state merging. However, Oppenheim~\cite{Opp08} proved that the reverse is also true: 
state redistribution can be achieved in the asymptotic i.i.d.~ setting by using coherent state merging as a primitive. In this paper we make use 
of this idea, and employ the bounds on the quantum communication cost and 
entanglement gain for one-shot coherent state merging derived in~\cite{DH11}, 
to obtain an upper bound on the quantum communication cost for one-shot state redistribution.

In Section~\ref{defns} we define the entropic quantities in terms 
of which our results, Theorem~\ref{thm:main} and Theorem~\ref{thm:main2}, are expressed, and state some of their relevant 
properties. 
In addition, we define the operational quantities of one-shot coherent state merging
which we employ in our proof of Theorem~\ref{thm:main}. In Section~\ref{sec:state_redist} we give a precise definition of the operational quantity that we study, namely, the quantum communication cost of $\eps$-error one-shot state redistribution and state our first theorem (Theorem~\ref{thm:main}) which consists
of an upper bound on this cost. In Section~\ref{sec:FQSW} we recall the protocol of coherent state merging, which we
use as a primitive in our proof of Theorem~\ref{thm:main}, which is given in Section~\ref{sec:proof}. The statement and proof of
our main result (Theorem~\ref{thm:main2}), which consists of an upper bound on the second order asymptotic expansion for the quantum communication cost of state redistribution, is given in Section~\ref{sec:second}.

\section{Notations and Definitions}\label{defns}
Let $\cP(\cH)$ denote the set of positive semi-definite operators acting on a finite-dimensional Hilbert space $\cH$, and let ${\cD}({\cH})\subset\cP(\cH)$ denote the set of density matrices (states) on $\cH$. Furthermore, let ${\cD}_{\leq}({\cH})$ denote the set of subnormalized states\footnote{Throughout this paper, we restrict our considerations to finite-dimensional Hilbert spaces and take the logarithm to base $2$.}.
For any given pure state $|\psi\rangle \in {\cH}$, we denote the projector $|\psi\rangle\langle\psi|$ simply as $\psi$. For $\omega_{AB} \in \cP(\cH_A \otimes \cH_B)$, let $\omega_{A} := \tr_B\omega_{AB}$ denote its restriction to the subsystem $A$. For $\rho, \sigma \in {\cD}({\cH})$,
the fidelity is defined as
$F(\rho, \sigma):= \tr\sqrt{\sqrt{\rho} \sigma \sqrt{\rho}}$. We use the same expression for fidelity when either one of $\rho$ or $\sigma$ is subnormalized. For simplicity, we denote a quantum operation (i.e., a completely positive trace-preserving (CPTP)
map) $\Lambda: \cD(\cH_A) \mapsto \cD(\cH_B)$ as $\Lambda: A \mapsto B$. The identity map is denoted as ${\rm{id}}$. A quantum operation on a bipartite system, shared between two distant parties (say, Alice and Bob), which consists of local operations on the two subsystems 
and quantum communication from Alice to Bob is said to be a (one-way) LOQC map.

The results in this paper involve various entropic quantities. The von Neumann entropy of a state $\rho_A\in\cD(\cH_A)$ is given by $H(A)_{\rho} = - \tr \rho_A \log \rho_A$. For a bipartite system, $\rho_{AB}$ the conditional entropy of the subsystem $A$ given $B$ is defined as $H(A|B) = H(\rho_{AB}) - H(\rho_B)$. For a tripartite state $\rho_{ABC}$, the conditional mutual information of the subsystems $A$ and $B$ given $C$ is defined as:
$$I(A;B|C) = H(B|C) - H(B|AC).$$

In addition to the above entropic quantities, we make use of the following generalized entropies~\cite{renner, KRS09} which arise naturally in one-shot quantum information theory:
\smallskip

\noindent
Let $\rho_{AB}\in\cD_{\leq}(\cH_{A}\otimes\cH_B)$. For a bipartite state $\rho_{AB}$, the min-entropy of $A$ conditioned on $B$ is defined as
\[ 
H_{\min}(A|B)_\rho=\max_{\sigma_B\in\cD(\cH_B)}\left[-D_{\max}(\rho_{AB}||I_A\otimes\sigma_B)\right],
\]
where for any $\rho\in\cD_{\leq}(\cH)$ and $\omega\in\cP(\cH)$, $D_{\max}(\rho||\omega)$ is the max-relative entropy~\cite{min-max}:
$$D_{\max}(\rho||\omega):= \inf\{\gamma:\rho\leq2^\gamma \omega\}.$$
For any $\eps\in (0, 1)$, a smooth version of these quantities are given by
\begin{align}\label{smoothmin}
D_{\max}^\eps(\rho||\omega) &:= \min_{\overline{\rho}\in\cB_\eps(\rho)}D_{\max}({\overline{\rho}}||\omega)\nonumber\\
H_{\min}^\eps(A|B)_\rho &:=\max_{\overline{\rho}_{AB}\in\cB_\eps(\rho_{AB})} H_{\min}(A|B)_{\overline{\rho}},
\end{align}
where for any state $\rho\in\cD(\cH)$, $\cB_\eps(\rho)$ denotes the $\eps$-ball around $\rho$ 
and is defined as
\[
\cB_\eps(\rho):=\{{\overline{\rho}}\in\cD_{\leq}(\cH):F^2({\overline{\rho}}, \rho)\geq1-\eps^2\}.
\]

The smooth  conditional max-entropy is given in terms of the smooth conditional min-entropy via the following duality relation \cite{TCR09, TCR10, KRS09}:

Let $\rho_{AB}\in\cD(\cH_{A}\otimes\cH_B)$ and let $\rho_{ABC}\in\cD(\cH_{A}\otimes\cH_B\otimes\cH_C)$ be an arbitrary purification of $\rho_{AB}$. Then for any $0\leq \eps\leq 1$,
\be\label{duality}
H_{\max}^\eps(A|C)_\rho := - H_{\min}^\eps(A|B)_\rho.
\ee

We also make use of the R\'enyi entropy of order zero, which for a state $\rho \in \cD(\cH)$ is defined as
\[
H_0(A)_\rho =\log ({\rm{rk}} \rho_A),
\]
where ${\rm{rk}} \rho_A$ denotes the rank of $\rho_A$. Its smooth version for any $\eps \in (0,1)$ is given by
\[H_0^\eps(A)_\rho = \min_{\overline{\rho}\in\cB_\eps(\rho)} H_{0}(A)_{\overline{\rho}}.\]
In order to obtain an upper bound on the second order asymptotic expansion for the quantum communication cost, we make use of the second order asymptotic expansion for the smooth max-relative entropy which was derived by Tomamichel and Hayashi in~\cite{TH13}: for any $\rho \in \cD(\cH)$ and $\sigma \in \cP(\cH)$ with ${\rm{supp}} \rho \subseteq {\rm{supp}} \sigma$, $\forall \, \eps \in (0,1)$:
\begin{align}\label{secondmax}
D_{\max}^\eps(\rho^{\otimes n}\|\sigma^{\otimes n}) &= nD(\rho\|\sigma) - \sqrt{n}\,\mathfrak{s}(\rho\|\sigma)\Phi^{-1}(\eps^2)+\cO(\log n),
\end{align}
where $D(\rho\|\sigma) := \tr\left(\rho \log \rho - \rho \log \sigma\right)$ is the quantum relative entropy, 
\be\label{eq:frak-s}
\sk(\rho\|\sigma):= \sqrt{V(\rho\|\sigma)}, \quad {\hbox{with}}\quad  V(\rho\|\sigma) :=  \tr\left[\rho(\log\rho-\log\sigma)^2\right]-D(\rho\|\sigma)^2,
\ee
being the {\em{quantum information variance}}, and $\Phi^{-1}(\eps) :=  \sup\lbrace x\in\mathbb{R}\mid \Phi(x)\leq \eps\rbrace$ is the inverse of the cumulative distribution function of a standard normal random variable.
\section{One-shot state redistribution}\label{sec:state_redist}
Our first result is an upper bound on the $\eps$-error quantum communication cost of state redistribution. It is given by Theorem~\ref{thm:main} below. Before stating it we need the following definition. 

\begin{definition}(One-shot state redistribution) \\
Consider a tripartite state  $\rho_{ABC}$  shared between two parties Alice and Bob, with the systems $A$ and $C$ being with Alice and the system $B$ being with Bob.  Let $\psi_{ABCR}$ denote a purification of $\rho_{ABC}$, with $R$ being the inaccessible, purifying reference system.
Let Alice and Bob have further registers $A_0$, $A_1$ and $B_0$, $B_1$, respectively. A {\em{one-shot $\eps$-error state redistribution}} protocol is then defined as a joint quantum operation $\cM : ACA_0 \otimes BB_0 \to 
CA_1 \otimes B_1{B^\prime}B$, 
which is one-way LOQC (with the quantum communication being from Alice to Bob) and 
such that 
\be 
F\left(\rho_{CA_1B_1{B^\prime} BR},\Phi^m_{A_1B_1} \otimes \psi_{C{B^\prime} BR}\right) \ge 1 - \eps,
\label{fid}
\ee
where $\rho_{CA_1B_1{B^\prime}BR} := \left(\cM \otimes {\rm{id}}_R\right)\left(\psi_{ABCR}\otimes \Phi^k_{A_0B_0}\right)$ and
$\Phi^k_{A_0B_0}$, $\Phi^m_{A_1B_1}$ are maximally entangled states of 
Schmidt rank $k$, $m$, respectively. Here, $B^\prime$
is a local
ancilla of Bob's of the same size as $A$. The quantum communication cost of the protocol, which we denote as $q^{(1)}_{\eps} (\rho_{ABC}, \Lambda)$, is  the minimum number of qubits that Alice needs to send to Bob for \reff{fid} to hold.
Moreover, the number $(\log k - \log m)$ 
is called the entanglement cost of the protocol. 
\end{definition}

The quantum communication cost of $\eps$-error one-shot state redistribution for a state $\rho_{ABC}$ is then defined as 
\be\label{qcc}
q^{(1)}_{\eps} (\rho_{ABC}):= \min_{\Lambda} q^{(1)}_{\eps} (\rho_{ABC}, \Lambda),
\ee
where the minimum is taken over all $\eps$-error one-shot state redistribution
protocols $\Lambda$.

In this paper we only focus on the quantum communication cost. Our main result is given by the following theorem.
\begin{theorem}\label{thm:main}
Fix $\eps\in (0,1)$. Then for any tripartite state $\rho_{ABC}$, there exists an $\eps$-error one-shot state redistribution protocol $\Lambda$, with quantum communication cost given by
\be
q^{(1)}_{\eps}(\rho_{ABC}, \Lambda) = \frac{1}{2}\left(H_{\max}^{\eps^\prime}(A|B)_{\psi}-H_{\min}^{\eps^\prime}(A|RB)_{\psi}\right) - 2\log\eps^{\prime},
\label{sr1}
\ee
where $\eps^{\prime} = \eps^2/(\sqrt{5} + 1)^2$, and $\psi_{AB}$ and $\psi_{ABR}$ are the reduced states of a purification $\psi_{ABCR}$ of the state $\rho_{ABC}$.

In particular, the RHS of \reff{sr1} provides an upper bound on the quantum communication cost $q^{(1)}_{\eps} (\rho_{ABC})$ defined by \reff{qcc}.
\end{theorem}
\section{One-shot coherent state merging: a primitive for one-shot state redistribution}\label{sec:FQSW}
The proof of Theorem~\ref{thm:main} employs a result on one-shot coherent state merging (or FQSW) proved in~\cite{DH11}, which is given by Theorem~\ref{thm:FQSW} below. Before stating it, we need 
to introduce the following definition.
\begin{definition} [One-shot coherent state merging or FQSW]

Consider a bipartite state  $\rho_{AB}$  shared between Alice and Bob, with the system $A$ being with Alice and the system $B$ being with Bob.  
Let $\psi_{ABR}$ denote its purification, with $R$ being the inaccessible, purifying reference system. 
We call a quantum operation ${\tilde{\cM}} : A\otimes B\to 
A_1 \otimes B_1{B^\prime}B$ {\em{one-shot $\eps$-error coherent state merging}} of $\rho_{AB}$ 
if it is one-way LOQC (with the quantum communication being from Alice to Bob) and
the state 
$\Omega_{A_1B_1{B^\prime}BR} := \left({\tilde{\cM}} \otimes {\rm{id}}_R\right)\psi_{ABR}$, is such that
\be 
F\left(\Omega_{A_1B_1{B^\prime} BR},\Phi^m_{A_1B_1} \otimes \Psi_{{B^\prime} BR}\right) \ge 1 - \eps,
\label{fid2}
\ee
where $\Phi^m_{A_1B_1}$ denotes a maximally entangled state of Schmidt rank $m$. Here, $B^\prime$
is a local
ancilla of Bob's of the same size as $A$. The number $\log m$ 
is called the {\em{entanglement gain}} of the protocol and denoted as 
${\te}^{(1)}_{\eps}(\rho_{AB}, {\tilde{\cM}})$. 
Let $\tq^{(1)}_{\eps} (\rho_{AB}, {\tilde{\cM}})$ denote the corresponding quantum communication cost, that is the minimum 
number of qubits that Alice needs to send to Bob for \reff{fid2} to hold.
\end{definition} 

\begin{theorem}[\cite{DH11}]
\label{thm:FQSW}
Fix $\eps\in (0,1)$. Then for any bipartite state $\rho_{AB}$, there exists an $\eps$-error one-shot coherent state merging protocol, ${\tilde{\cM}}$, with entanglement gain and quantum communication cost respectively given by\footnote{The choice of the smoothing parameter $\eps^\prime$ here is different from that in~\cite{DH11} because here we choose the figure of merit for the coherent state merging protocol to be given by the fidelity, whereas in~\cite{DH11} we chose it to be the trace distance.},
\begin{align}
{\te}^{(1)}_{\eps} (\rho_{AB}, {\tilde{\cM}})&= \frac{1}{2}\left[H^{{\eps^\prime}}_0(A)_\psi+H_{\min}^{{\eps^\prime}}(A|R)_{\psi}\right]+\log\eps^{\prime}\label{five}\\
\tq^{(1)}_\eps (\rho_{AB}, {\tilde{\cM}})&= \frac{1}{2}\left[H_{ 0}^{{\eps^\prime}}(A)_\psi- H_{\min}^{{\eps^\prime}}(A|R)_\psi\right]-\log\eps^{\prime}\label{six}
\end{align}
where $\eps^{\prime} = \eps^2/(\sqrt{5} + 1)^2$, and $\psi_{A}$ and $\psi_{AR}$ are the reduced states of a purification $\psi_{ABR}$ of the state $\rho_{AB}$.
\end{theorem}
\smallskip

\noindent
{\bf{Remark:}} The proof of the above theorem, given in \cite{DH11}, relies on a decoupling argument and ensures the existence of a unitary operator $U$ and an isometry $V$,
such that $\eps$-error one-shot coherent state merging is achieved if (i) Alice acts on the state of her system $A$ with $U$, (ii) sends $\tq^{(1)}_\eps$ qubits to Bob, and (iii) Bob acts 
on the composite state of the qubits that he receives from Alice and the state of his system $B$ by the isometry $V$. 

\section{Proof of Theorem~\ref{thm:main}}\label{sec:proof}
For any tripartite state $\rho_{ABC}$, an expression for the quantum communication cost of an $\eps$-error one-shot state redistribution protocol, for any fixed $\eps \in (0,1)$,  can be obtained by a direct application of one-shot coherent state merging, if we simply
consider Alice to transfer the state of her system $A$ to Bob, without exploiting the additional system $C$ which is in her possession. In this case,
we can consider $C$ to be part of the reference system. From eq.\reff{six} of Theorem~\ref{thm:FQSW} we then infer that state redistribution can be achieved
by the transfer of the following number of qubits from Alice to Bob:
\be
\Delta q = \frac{1}{2}[H_{ 0}^{{\eps^\prime}}(A)_\psi- H_{\min}^{{\eps^\prime}}(A|CR)_\psi]-\log\eps^{\prime},
\label{naive}
\ee
where $\eps^\prime$ is as stated in Theorem~\ref{thm:main}. 

\begin{figure}[htbp]
\centering
  \fbox{\includegraphics[width=8cm]{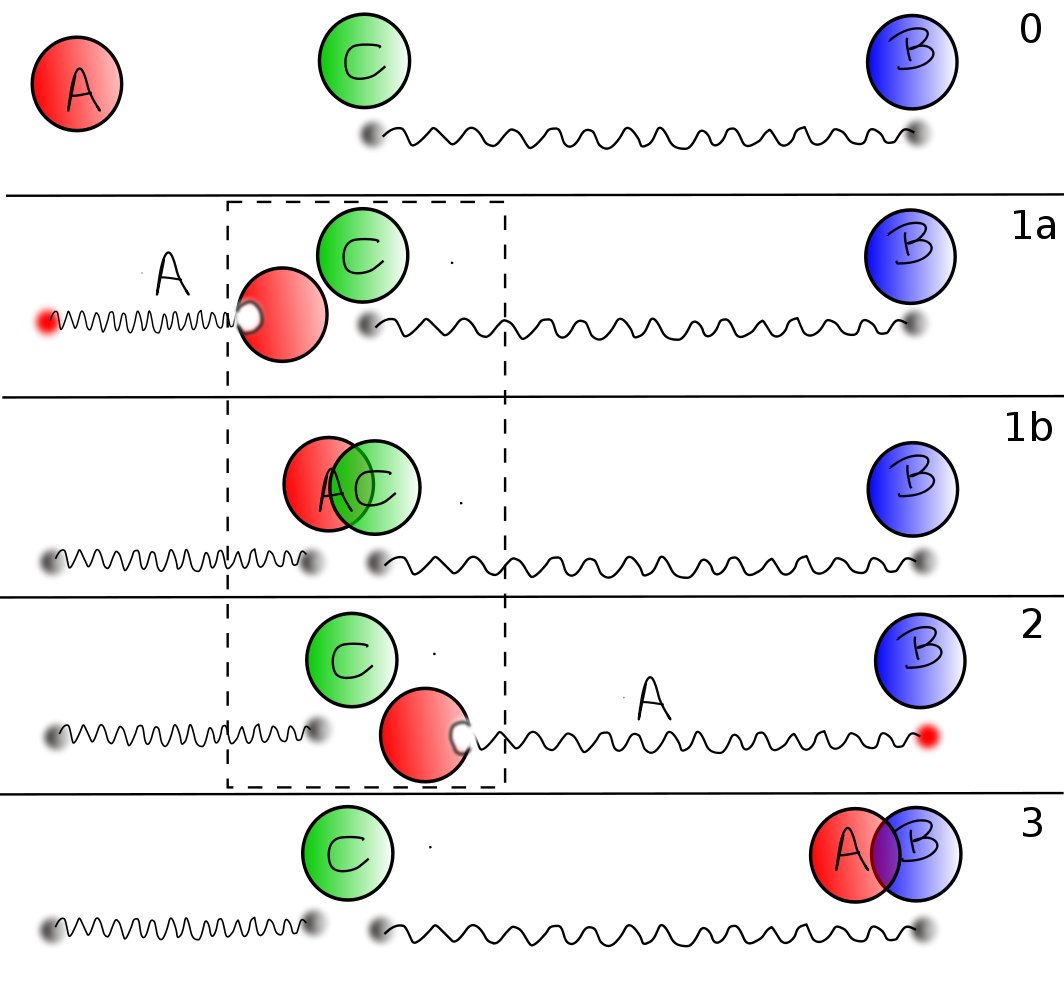}}
   \caption{\em{The $\eps$-error one-shot state redistribution protocol using
one-shot coherent state merging and ebit repackaging.
Shares of the state $\rho_{ABC}$ are represented by circles, while shared
entanglement is represented by wiggly lines.  The protocol of ebit
repackaging is
contained in the dashed rectangle.
Due to the ebit repackaging, 
${\te}^{(1)}_{\eps} (\rho_{AB}, {\tilde{\cM}})$ qubits (highlighted in
red and given by \reff{five} of Theorem~\ref{thm:FQSW}) are effectively
sent to Bob without being physically transferred.
Here ${\tilde{\cM}}$ denotes the coherent state merging protocol from Alice
to Charlie.}}
  \label{fig:protocol}
\end{figure}

However, one-shot state redistribution can be achieved at a lower quantum communication cost than that given by \reff{naive} above. A simple way to see this is by employing the one-shot version
of a novel construction which was introduced by Oppenheim~\cite{Opp08} in the asymptotic i.i.d.~setting. In it the system $C$ plays the role of a coherent relay as explained below (see also Figure \ref{fig:protocol}).
For this construction, it is convenient to split the two-party protocol between Alice and Bob into a three-party protocol, by considering the system $C$ to be 
in the possession of a third party (say, Charlie). The construction is implemented through the following steps:
\smallskip

\noindent
{\em{Step 1.}} {\em{One-shot $\eps$-error coherent state merging from Alice to Charlie}}: Let us denote this protocol by ${\tilde{\cM}}$. It results in the transfer of Alice's state $\rho_A$ to Charlie (with an error 
of atmost $\eps$),
with the simultaneous generation of entanglement between them. By eq.\reff{five} of Theorem~\ref{thm:FQSW}, the number of ebits of entanglement generated is given by 
\be
 \Delta e:= \frac{1}{2}\left[H^{{\eps^\prime}}_0(A)_\psi+H_{\min}^{{\eps^\prime}}(A|BR)_{\psi}\right]+\log\eps^{\prime}.
\label{ent-gen}
\ee
As explained in \cite{Opp08}, from the remark given after Theorem~\reff{thm:FQSW} it follows that this step can itself be broken down into two steps: {{(i)}} Alice applies a unitary transformation denoted by a unitary operator $U$ (say)
on her system $A$ and sends the required number of qubits (needed to implement a one-shot $\eps$-error coherent state merging protocol ${\tilde{\cM}}$) 
to Charlie. This number is given by the right hand side of eq.\reff{six} of Theorem~\ref{thm:FQSW}, with the replacement of $R$ by $BR$. This is because
in this case $BR$ play the role of the reference. {{(ii)}} 

 Charlie then does the corresponding decoding isometry $V$ (say) on the composite state
of the qubits that he receives from Alice and the system $C$ in his possession. After applying $V$,
the resulting output will be ${\te}^{(1)}_{\eps} (\rho_{AB}, {\tilde{\cM}})$ ebits, shared between him and Alice, and a remaining subsytem $S$ in Charlie's possession.
\smallskip

\noindent
{\em{Step 2.}} {\em{Ebit repackaging:}} Charlie sets aside {{his share of the ebits}} that were generated from the previous step and
replaces them by those that he shared with Bob at the start of the protocol. He then applies $V^\dagger$ to the joint state of the latter 
and $S$.

\smallskip

Note that the above steps effectively result in the transfer of $\Delta e$ qubits from Alice to Bob. Hence, instead of sending $\Delta q$ qubits (given by \reff{naive}),
Alice (or, in this three-party description, Charlie) only needs to physically send $\left(\Delta q - \Delta e\right)$ qubits to Bob, in order to achieve $\eps$-error one-shot state redistribution. From \reff{naive} and
\reff{ent-gen} we then infer that there exists an $\eps$-error one-shot state redistribution protocol, $\Lambda$, with quantum communication cost:
\begin{align}
q^{(1)}_\eps(\rho_{ABC}, \Lambda) &=\Delta q - \Delta e \nonumber\\
&=  \frac{1}{2}\left(- H_{\min}^{{\eps^\prime}}(A|CR)_\psi - H_{\min}^{{\eps^\prime}}(A|BR)_{\psi}\right) - 2\log\eps^{\prime},\nonumber\\
&=  \frac{1}{2}\left(H_{\max}^{{\eps^\prime}}(A|B)_\psi  - H_{\min}^{{\eps^\prime}}(A|BR)_{\psi}\right) - 2\log\eps^{\prime},
\end{align}
where the last line follows from the duality relation \reff{duality}, since $\psi_{ABCR}$ is a pure state. This completes the proof of Theorem~\ref{thm:main}.

\section{Second order asymptotics}\label{sec:second}
Consider the situation in which Alice and Bob share $n$ identical copies of the state $\rho_{ABC}$. In this case, it follows from Theorem~\ref{thm:main} that an upper bound on the quantum communication cost, $q^{(1)}_{\eps}(\rho_{ABC}^{\otimes n})$, for state redistribution is given by the following:
\begin{align}\label{second1}
q^{(1)}_{\eps}(\rho_{ABC}^{\otimes n}) &\le \frac{1}{2}\left[H_{\max}^{\eps^\prime}(A_n|B_n)_{\psi}-H_{\min}^{\eps^\prime}(A_n|R_nB_n)_{\psi}\right] - 2\log\eps^{\prime},\nonumber\\
&=  \frac{1}{2}\left[-H_{\min}^{\eps^\prime}(A_n|C_nR_n)_{\psi}-H_{\min}^{\eps^\prime}(A_n|R_nB_n)_{\psi}\right] - 2\log\eps^{\prime},\nonumber\\
& = \frac{1}{2}[\min_{\sigma_{C_nR_n}}D_{\max}^{\eps^\prime}(\psi_{A_nC_nR_n}||I_{A_n} \otimes \sigma_{C_nR_n})
\nonumber\\
& \quad +  \min_{\omega_{B_nR_n}} D_{\max}^{\eps^\prime}(\psi_{A_n B_nR_n} || I_{A_n} \otimes \omega_{B_nR_n})] - 2\log\eps^{\prime}\nonumber\\
 & \le \frac{1}{2}\bigl[ \min_{\sigma_{CR}} D_{\max}^{\eps^\prime}(\psi_{ACR}^{\otimes n}||I_{A}^{\otimes n} \otimes \sigma_{CR}^{\otimes n})\nonumber\\
& \quad +  \min_{\omega_{BR}} D_{\max}^{\eps^\prime}(\psi_{ABR}^{\otimes n}||I_{A}^{\otimes n} \otimes \omega_{BR}^{\otimes n})\bigr] - 2\log\eps^{\prime}.
\end{align}
where $\psi_{A_nB_nC_nR_n} \equiv \psi_{ABCR}^{\otimes n}$, with
$\psi_{ABCR}$ being a purification of $\rho_{ABC}$. The first equality follows from the duality relation \reff{duality}, the second equality follows from the definition 
\reff{smoothmin} of the smooth conditional min-entropy, the second inequality follows from the restriction of the minimization to a smaller set, and the fact that the reduced states of $\psi_{A_nB_nC_nR_n}$ are tensor-power states. The minimizations in the above equation are all over (normalized) states.

We now employ the second order asymptotic expansion of the max-relative entropy, given by \reff{secondmax}, which we recall here for convenience:
$$D_{\max}^\eps(\rho^{\otimes n}\|\sigma^{\otimes n}) = nD(\rho\|\sigma) - \sqrt{n}\,\mathfrak{s}(\rho\|\sigma)\Phi^{-1}(\eps^2)+\cO(\log n).$$
Note that for $\eps \in (0,\frac{1}{2})$, $\Phi^{-1}(\eps)<0$ and hence the second term on the right hand side of the above equation is positive. 

Substituting the 
above expansion for the smooth max-relative entropies, with the smoothing parameter given by $\eps^\prime = \eps^2/(\sqrt{5} + 1)^2$, in the last line of \reff{second1}, we obtain: for any $\eps \in (0, 1/2)$,
\begin{align}
q^{(1)}_{\eps}(\rho_{ABC}^{\otimes n}) & \le n\Bigl[ \frac{1}{2}\bigl[\min_{\sigma_{CR}} D(\psi_{ACR}||I_{A} \otimes \sigma_{CR})+  \min_{\omega_{BR}} D(\psi_{ABR}||I_{A} \otimes \omega_{BR})\bigr]\Bigr]\nonumber\\
& \quad  + \frac{1}{2} \min_{\sigma_{CR}} \Bigl[\bigl(- \sqrt{n} \Phi^{-1}({\eps^\prime}^2)\bigr)\mathfrak{s}(\psi_{ACR}||I_{A} \otimes \sigma_{CR}) \Bigr]\nonumber\\
&\quad + \frac{1}{2} \min_{\omega_{BR}} \Bigl[\bigl(- \sqrt{n} \Phi^{-1}({\eps^\prime}^2)\bigr) \mathfrak{s}(\psi_{ABR}||I_{A} \otimes \omega_{BR})\Bigr]  + \cO(\log n),\nonumber\\
&\le n\bigl[ \frac{1}{2} I(A;R|B)_\psi\bigr]
- \sqrt{n} \Phi^{-1}({\eps^\prime}^2) \Bigl[\frac{1}{2}\bigl[\mathfrak{s}(\psi_{ACR}||I_{A} \otimes \psi_{CR})
+  \mathfrak{s}(\psi_{ABR}||I_{A} \otimes \psi_{BR})\bigr]\Bigr] \nonumber\\
& \qquad + \cO(\log n),
\label{last2}
\end{align}
To arrive at the last line of \reff{last2}, we used the following facts: \\
$(i)$ $\eps^\prime < \eps$ and hence $\Phi^{-1}({\eps^\prime}^2)<0$;\\
$(ii)$ for any bipartite state $\rho_{AB}$, $\min_{\sigma_B \in \cD(\cH_B)} D(\rho_{AB}||I_A \otimes \sigma_B) = D(\rho_{AB}||I_A \otimes \rho_B),$ which simply follows from the fact that the relative entropy of two states is non-negative (see e.g.~Lemma 6 of~\cite{BD10})\\
$(iii)$ For a pure state $\psi_{ABCR}$, $H(A|CR)_\psi = - H(A|B)_\psi$, where $\psi_{ACR}$ and $\psi_{ABR}$ are the reduced states of $\psi_{ABCR}$.\\
$(iv)$ $I(A;R|B) = H(A|B) - H(A|BR).$
\smallskip

\noindent
Thus we have proved the following theorem, which constitutes our main result:
\begin{theorem}\label{thm:main2}
Fix $\eps \in (0, 1/2)$. Then for any tripartite state $\rho_{ABC}$, an upper bound on the second order asymptotic expansion for the quantum communication cost of achieving
state redistribution with an error of at most $\eps$, is given by
\begin{align}
&n\bigl[ \frac{1}{2} I(A;R|B)_\psi\bigr]
- \sqrt{n} \Phi^{-1}({\eps^\prime}^2) \Bigl[\frac{1}{2}\bigl[\mathfrak{s}(\psi_{ACR}||I_{A} \otimes \psi_{CR})
+  \mathfrak{s}(\psi_{ABR}||I_{A} \otimes \psi_{BR})\bigr]\Bigr] + \cO(\log n),
\end{align}
where $\eps^{\prime} = \eps^2/(\sqrt{5} + 1)^2$, and $\mathfrak{s}(\cdot ||\cdot )$, defined by \reff{eq:frak-s}, denotes the square root of the quantum information variance.
\end{theorem}

As a corollary of this theorem we recover the following result of Devetak and Yard~\cite{YD09}
stated earlier: {\em{in the asymptotic i.i.d.~setting, state redistribution for a tripartite state $\rho_{ABC}$ can be achieved if Alice sends qubits at a rate $(1/2) I(A;R|B)$ to Bob.}} 
This immediately follows from Theorem~\ref{thm:main2} since the quantum communication cost $Q$ in the asymptotic i.i.d.~setting can be expressed in 
terms of  $q^{(1)}_{\eps} (\rho_{ABC}^{\otimes n})$ as follows:
\be
Q \equiv Q(\rho_{ABC}) =  \lim_{\eps\to 0} \lim_{n\to \infty}\frac{1}{n} q^{(1)}_{\eps}(\rho_{ABC}^{\otimes n}).
\ee

\section*{Acknowledgements} We had stated the one-shot result given in Theorem~\ref{thm:main} in an earlier paper~\cite{DH11}. We are grateful to Mario Berta, Matthias Christandl and Dave Touchette for getting us to finally write up its proof. We are also grateful to Felix Leditzky for his comments on a first draft of this paper.



\begin{thebibliography}{~~} \label{refs}


\bibitem{DY06}  I.~Devetak and J.~Yard, The operational meaning of quantum conditional information,  arXiv:quant-ph/0612050.

\bibitem{WDHW13} M.~M.~Wilde, N.~Datta, M-H.~Hsieh, A.~Winter, Quantum rate distortion coding with auxiliary resources,
IEEE Transactions on Information Theory, vol.~59, no.~10, pp.~6755-6773 
(October 2013).

\bibitem{DL09} Z.~Luo and I.~Devetak, Channel simulation with quantum side information, IEEE Transactions on Information Theory, vol.~55, no.~3, pp.~1331–1342, March 2009, arXiv:quant-ph/0611008.
%
\bibitem{YD09} J.~Yard and I.~Devetak, Optimal quantum source coding with quantum
side information at the encoder and decoder, IEEE Transactions in
Information Theory, vol.~55, no.~11, pp.~5339–5351, November 2009, arXiv:0706.2907.

\bibitem{DY08} I.~Devetak and J.~Yard, Exact cost of redistributing multipartite quantum states, Physical Review Letters, vol.~100, no.~23, p.~230501, 
June 2008.

\bibitem{FQSW} A.~Abeyesinghe, I.~Devetak, P.~Hayden, and A.~Winter, The mother of all protocols: restructing
quantum information's family tree, Proceedings of the Royal Society A, vol.~465,
pp.~2537-2563, 2009.

\bibitem{Opp08} J.~Oppenheim, State redistribution as merging: introducing the coherent relay, arXiv:0805.1065, (2008).

\bibitem{DH11} N.~Datta and M.-H.~Hsieh, The apex of the family tree of protocols: optimal rates and resource inequalities, New Journal of Physics, vol.~ 13, p.~093042 (2011).

 \bibitem{renner} R.~Renner, {{Security of quantum key distribution}},
PhD thesis, {{arXiv:quant-ph/0512258}}.

\bibitem{marco} M.~Tomamichel, A Framework for Non-Asymptotic Quantum Information Theory,
{PhD Thesis, Department of Physics, ETH Zurich, arXiv:1203.2142.}

\bibitem{KRS09} R.~K\"onig, R.~Renner, and C.~Schaffner, The 
operational meaning of min- and max-entropy, Transactions on Information Theory, vol.~55, pp.4337-4347 (2009).

\bibitem{min-max} N.~Datta, Min- and Max- Relative Entropies and a New Entanglement Monotone, IEEE Transactions on Information Theory, vol.~55, p.~2816-2826 (2009).

\bibitem{TH13} M.~Tomamichel and M.~Hayashi, A hierarchy of information quantities for finite block
length analysis of quantum tasks, IEEE Transactions on Information Theory vol.~59  pp.~7693-7710 (2013).

\bibitem{TCR09} M.~Tomamichel, R.~Colbeck, and R.~Renner, 
Fully Asymptotic Equipartition Property, IEEE Transactions on Information Theory, vol.~55, pp.~5840-5847 (2009).

\bibitem{TCR10} M.~Tomamichel, R.~Colbeck, and R.~Renner, Duality between smooth min- and max-entropies,
IEEE Transactions on Information Theory, vol.~56, pp.~4674-4681 (2010).

\bibitem{BD10} F.~Buscemi and N.~Datta, The quantum capacity of channels with arbitrarily correlated noise,
IEEE Transactions on Information Theory, vol.~56, pp.~1447-1460 (2010).

\bibitem{BCT} M.~Berta, M.~Christandl and D.~Touchette, Smooth Entropy Bounds on One-Shot
Quantum State Redistribution. To appear on the arXiv simultaneously with this
paper.


\end{thebibliography}
\end{document}